\documentclass[a4paper]{jpconf}
\usepackage{graphicx}
\usepackage{amsmath}
\usepackage{comment}
\usepackage[toc]{appendix}
\usepackage{tabularx}
\usepackage{epsfig}
\usepackage{amssymb}
\usepackage{graphicx}
\usepackage{multirow}
\newcolumntype{R}[1]{>{\raggedleft\arraybackslash}p{#1}}

\newcommand{\aap}{Astron. \& Astrophys.}

\newcommand{\mnras}{MNRAS}

\newcommand{\prd}{Phys. Rev. D}

\begin{document}

\title{High energy neutrino and gamma-ray emissions from the jets of M33 X-7 microquasar}

\author{D A Papadopoulos$^1$, Th V Papavasileiou$^{1,2}$ 
and T S Kosmas$^{1}$}

\address{$^1$ Division of Theoretical Physics, University of Ioannina, GR-45110 Ioannina, Greece}
\address{$^2$ Department of Informatics, 
University of Western Macedonia, GR-52100 Kastoria, Greece}

\ead{mitsospapad@hotmail.com, theodora836@gmail.com, hkosmas@uoi.gr}


\begin{abstract}
In this work, after testing the reliability of our algorithms through numerical simulations on 
the well-studied SS 433 Galactic microquasar, 
we focus on neutrino and $\gamma$-ray emissions from the extragalactic 
M33 X-7 system. This is a recently discovered X-ray binary system located in the neighbouring galaxy 
Messier 33 which has not yet been modelled in detail. The neutrino and $\gamma$-ray energy spectra,
produced from the magnetized astrophysical jet of M33 X-7, in the context of our method are assumed
to originate from the decay (and scattering) processes taking place among the secondary particles
produced assuming that, first, hot (relativistic) protons of the jet scatter on thermal ones (p-p 
interaction mechanism).
\vskip 1em
\hspace{-2.5 em}

\end{abstract}



\section{Introduction}

During the last few decades, collimated outflows have been observed to emerge from a wide variety 
of astrophysical objects. Among these objects, the class of microquasars (MQs) and the X-ray binary 
systems posses prominent positions \cite{Smponias_tsk_2011,Smponias_tsk_2014}. These systems consist 
of a compact object at the center (a stellar 
mass black hole or a neutron star) and a companion (donor) main sequence star. Due to the strong
gravitational field of the compact object, mass from the companion star is accreted onto its 
equatorial region forming an accretion disc. Such systems constitute excellent laboratories for 
investigating astrophysical flows in our present study \cite{RomeroReview, Reynoso08,Reynoso09}. 
They are usually treated as magnetohydrodymamical flows 
emanating from the vicinity of the compact object, the stellar mass black hole.
We assume spinning black holes with masses up to few tens (30-50) of the Sun mass 
\cite{Smponias_tsk_2014,RomeroReview,Reynoso08}.

From the observed characteristics of MQs we came to the conclusion that they share a lot of similarities 
in their physical properties with the class of Active Galactic Nuclei (AGN) even though the latter 
are enormously different in scale compared to microquasars. In this work we restrict ourselves
to magnetized astrophysical outflows characterized by the hadronic content in their jets. We
concentrate on numerical simulations of their $\gamma$-ray and neutrino emissions \cite{PapadD,PapOKos}.

In the present calculations, after fixing our model parameters and testing our algorithms on the reproducibility
of some well known properties of the well-studied SS 433 Galactic microquasar \cite{Ody-Smpon-2015,Ody-Smpon-2017,Ody-Smpon-2018}, we performed detailed simulations for the Galactic Cyg X-1 system and
the extragalactic M33 X-7 \cite{PietschWC}. The latter system is a recently discovered X-ray binary system located
in the neighboring galaxy Messier 33 \cite{PietschWC}. The neutrino and $\gamma$-ray emissivities (spectra), that
may be obtained in the context of our method originate from the decay (and scattering) processes
taking place among the secondary particle produced assuming that non-thermal (relativistic) protons of the 
jet (a small portion of them equal to about $0.1\%$) scatter on thermal ones (p-p scattering mechanism) \cite{Reynoso08}.   

\section{Hadronic Mechanisms for astrophysical outflows}

In the models considered in this work, an accretion disk is present around the compact object, 
and a fraction of the accreted material is expelled in two oppositely directed jets
\cite{Smponias_tsk_2014}. We assume conical jets with a half-opening angle $\xi=7^{\circ}$ for the M33 X-7 and radius $r (z) = z tan \xi$, where the injection point is at a distance $z_0$ from the compact object. When $z=z_0$ the radius of the jet is given by $r_0 = z_0 tan \xi$. Assuming an initial jet radius $r_0=r(z_0)=5R_{sch}$, where $R_{sch}=2GM_{BH}/c^2$, we find that the injection point is at 
$z_0= r_0/tan \xi \simeq 1.9 \times 10^8$ cm for the M33 X-7 ($M_{BH}=15.65 M_{\odot}$). 

In Table 1 we tabulate the parameters of the model for the system M33 X-7.

\begin{table} [ht]
\caption{Parameters of the model for M33 X-7} 
\centering
\begin{tabular}{|c|c|c|}
\hline
Parameter & Symbol & Value \\ [0.20ex]
\hline 
jet's launching point & $z_0$ & $1.9\times 10^8$ cm  \\
extent of acceleration region & $z_{max}$ & $5z_0$  \\
jet's bulk Lorentz factor & $\Gamma_b$ & 1.66  \\
jet's half-opening angle & $\xi$ & $7^{\circ}$  \\
viewing angle & $\theta$ & $74.6^{\circ}$  \\
\hline 
\end{tabular}
\end{table}

\subsection{The p-p Collision Mechanism}

The collision of relativistic protons with the cold ones (p-p collision mechanism) inside the jet, 
produces pions, kaons, eta particles, etc., through the following reactions 
\begin{equation}
\begin{split}
p+p & \rightarrow p+p+ a \pi^0 +b(\pi^+ + \pi^-)  \\ 
p+p & \rightarrow p+n+ \pi^+ + a \pi^0 +b(\pi^+ + \pi^-)
\end{split}
\end{equation}
where $a$ and $b$ denote the pion multiplicities \cite{RomeroReview}.
Charged pions ($\pi^\pm$), afterwards, decay to charged leptons (electrons or positrons and muons) 
and neutrinos as
\begin{equation}
\begin{split}
\pi^+ \rightarrow \mu^+ +\nu_{\mu}, \quad \quad \pi^+ \rightarrow e^+ + \nu_e \\
\pi^- \rightarrow \mu^- + \bar{\nu_{\mu}}, \quad \quad \pi^- \rightarrow e^- + \bar{\nu_e}
\end{split}
\end{equation}
Furthermore, muons also decay giving neutrinos and electrons (or positrons) as
\begin{equation}
\mu^+\rightarrow e^+ +\nu_e +\bar{\nu_{\mu}} , \quad \quad \mu^-\rightarrow e^- +\bar{\nu_e} + \nu_{\mu} .
\end{equation}
Me mention that neutral pions decays give $\gamma$-rays according to the reactions
\begin{equation}
\pi^0 \rightarrow \gamma + \gamma \, , \quad \quad \pi^0 \rightarrow \gamma + e^- + e^+ 
\end{equation}

In our present study we consider both types of pion decays \cite{PapadD,PapOKos}

\section{ Cooling rates for some important processes }

The rate of p-p collisions between the relativistic protons with the cold ones is 
given by
\begin{equation}
t_{pp}^{-1} = n(z) \sigma_{pp}^{inel} (E_p) K_{p}
\end{equation}
where the inelasticity coefficient is $K_{p} \approx 1/2$, the corresponding cross 
section for inelastic $p-p$ interactions is given in Ref. \cite{Kelner}. $n(z)$ is 
the density of cold particles in the jet at a distance z from the black hole given 
by
\begin{equation}
n(z)=\frac{(1-q_{rel})}{\Gamma m_pc^2\pi r_j^2 \upsilon_b}L_k
\end{equation}
Charged particles of mass $m$ and energy $E=\gamma m c^2$, emit synchrotron 
radiation at a rate
\begin{equation}
t_{sync}^{-1} = \frac{4}{3}{\left(\frac{m_e}{m}\right)}^3 \frac{\gamma \sigma_T B^2}{m_e c 8\pi}
\end{equation}
Finally, because the jet is expanding with a velocity $\upsilon_b tan \xi$) the 
adiabatic cooling rate is \cite{Bosch}
\begin{equation}
t_{ad}^{-1}=\frac{2}{3} \frac{\upsilon_b}{z}
\end{equation}

\section{Method of calculating particle distributions}

In order to calculate the neutrino and gamma-ray emissivities, we need, first, to
calculate the distributions of protons, pions and muons. For these calculations we 
use a code written in the C programming language, mainly following the assumptions 
of Refs. \cite{Reynoso08,Reynoso09}. In the one-zone approximation \cite{Ody-Smpon-2018},
the particle distributions are independent of time (steady state approximation) and 
can be obtained from the solutions of the following transport equations
\begin{equation}
\frac{\partial}{\partial E}
\begin{pmatrix} N_p(E,z)b_p(E,z) \\ N_{\pi}(E,z)b_{\pi}(E,z) \\ N_{\mu}(E,z)b_{\mu}(E,z) \end{pmatrix} + \begin{pmatrix} t_{esc}^{-1} N_p(E,z) \\ t_{\pi}^{-1} N_{\pi}(E,z) \\ t_{\mu}^{-1} N_{\mu}(E,z) \end{pmatrix} = \begin{pmatrix} Q_p(E,z) \\ Q_{\pi}(E,z) \\ Q_{\mu}(E,z) \end{pmatrix}
\end{equation}
where
\begin{equation}
t_{esc}^{-1} \approx \frac{c}{z_{max}-z}, \qquad   b(E,z)= \frac{dE}{dt} = -Et_{loss}^{-1} (E,z) \, .
\end{equation}
The first equation gives the escape rate and the second the energy loss rate. The 
energy loss rate for each particle is given by
\begin{equation}
b_p(E,z)=-E(t_{syn}^{-1} + t_{ad}^{-1} + t_{pp}^{-1} ), \quad b_{\pi}(E,z)=-E(t_{syn}^{-1} + t_{ad}^{-1} + t_{\pi p}^{-1} ), \quad b_{\mu}(E,z)=-E(t_{syn}^{-1} + t_{ad}^{-1})
\end{equation}
For the $\pi p$ interactions we consider
\begin{equation}
t_{\pi p}^{-1}(E,z) \approx 0.5 \ n(z)c \sigma_{\pi p}^{inel}(E_p)
\end{equation}
with $\sigma_{\pi p}(E) \approx 2 \sigma_{pp}^{inel}(E)/3$, which is based on the fact
that protons are made of three valence quarks, while the pions by only two quarks \cite{Gaisser}.

\subsection{Proton distribution}

The solution of the transport equation for protons is written as
\begin{equation}
N_p(E,z)=\frac{1}{\mid b_p(E) \mid}\int_{E_p}^{E_p^{max}} Q_p(E',z)e^{-t_{esc}^{-1}\tau (E,E')}dE', \qquad \tau (E,E')=\int_{E}^{E'} \frac{dE"}{\mid b(E")\mid}
\end{equation}
The quantity $Q_p(E,z)$ corresponds to the injection function of protons \cite{Reynoso08}, where $\Gamma_b$ is the bulk Lorentz factor of the jet. The normalization constant $Q_0$ 
is obtained by specifying the power in the relativistic protons \cite{Reynoso08}. The minimum
energy of protons is $E_p^{min}=1.2$ GeV and the maximum energy is assumed to be
$E_p^{max}=10^7$ GeV.

\subsection{Pion and muon distributions}

The steady state pion and muon distributions obey the transport equation with the replacement $t_{esc}^{-1} \rightarrow t_{\pi}^{-1}(E,z)$, for pions, and $t_{esc}^{-1} \rightarrow t_{\mu}^{-1}(E,z)$, for muons. The corresponding solutions are
\begin{equation}
\begin{split}
N_{\pi}(E,z) & =\frac{1}{\mid b_{\pi}(E)\mid}\int_{E}^{E_{max}}Q_{\pi}(E',z)e^{-\tau_{\pi}}dE'     \\
N_{\mu}(E,z) & =\frac{1}{\mid b_{\mu}(E)\mid}\int_{E}^{E_{max}}Q_{\mu}(E',z)e^{-\tau_{\mu}}dE' \, .
\end{split}
\end{equation}
The rate of decay and escape (for pions or muons) is:
\begin{equation}
t_{\pi,\mu}^{-1}(E,z)=t_{esc}^{-1}(z)+t_{dec}^{-1}(E)
\end{equation}
where $t_{dec}^{-1}= {[(2.6\times 10^{-8}) \gamma_{\pi}]}^{-1} \ s^{-1}$, for pions, 
and $t_{dec}^{-1}={[(2.2\times 10^{-6}) \gamma_{\mu}]}^{-1} \ s^{-1}$, for muons.

The solution solution of the transport equation that corresponds to no energy-losses 
takes the simple form
\begin{equation}
\begin{pmatrix} N_{\pi,0}(E,z) \\ N_{\mu,0}(E,z) \end{pmatrix} = \begin{pmatrix}
\frac{Q_{\pi}(E,z)}{t_{\pi}^{-1}(E,z)} \\ \frac{Q_{\mu}(E,z)}{t_{\mu}^{-1}(E,z)}
\end{pmatrix}
\end{equation}
In Ref. \cite{PapadD,PapOKos} we have calculated the distributions of Eqs. (14) and (16).

\subsubsection{Pion injection}

The injection function of pions, produced by p-p interactions, is given by
\begin{equation}
Q_{\pi}(E,z)=n(z)c\int_{\frac{E}{E_p^{max}}}^{1}N_p\left(\frac{E}{x},z\right)F_{\pi}\left(x,\frac{E}{x}\right)\sigma_{pp}^{inel}\left(\frac{E}{x}\right)\frac{dx}{x},
\end{equation}
where $x=E/E_p$ and $F_{\pi}$ denotes the distribution of pions produced per 
$p-p$ collision \cite{Kelner}.

\subsubsection{Muon injection}

In order to take the muon energy loss into account \cite{Lipari07}, it is necessary to consider the production of left handed and  right handed muons separately because they have different decay spectra. The injection functions of the left handed and right handed muons are \cite{Lipari07}:
\begin{equation}
\begin{split}
Q_{\mu_L^-,\mu_R^+}(E_{\mu},z) & = \int_{E_{\mu}}^{E^{max}} d E_{\pi} t_{\pi,dec}^{-1} (E_{\pi}) N_{\pi}(E_{\pi},z) \frac{r_{\pi}(1-x)}{E_{\pi}x{(1-r_{\pi})}^2} \Theta(x-r_{\pi}), \\          
Q_{\mu_R^-,\mu_L^+}(E_{\mu},z) & = \int_{E_{\mu}}^{E^{max}} d E_{\pi} t_{\pi,dec}^{-1} (E_{\pi}) N_{\pi}(E_{\pi},z) \frac{(x-r_{\pi})}{E_{\pi}x{(1-r_{\pi})}^2}\Theta(x-r_{\pi}),
\end{split}
\end{equation}
with $x = E_{\mu}/E_{\pi}$ and $r_{\pi} = {(m_{\mu}/m_{\pi})}^2$.\\

\subsection{Gamma-ray emission from the p-p interaction mechanism}

The p-p collision mechanism in the jets, produce secondary $\gamma$-rays. 
For $E_{\gamma}\geq 100$ $GeV$, we consider the $\gamma$-ray emissivity at 
a height $z$ along the jets (in units $GeV^{-1} s^{-1}$) as
\begin{equation}
Q_{\gamma}= \int_{\frac{E_{\gamma}}{E_p^{max}}}^1 \sigma_{pp}^{inel}\left(\frac{E_{\gamma}}{x}\right) c N_p\left(\frac{E_{\gamma}}{x},z\right) F_{\gamma}\left(x,\frac{E_{\gamma}}{x}\right)\frac{dx}{x}
\end{equation}
The spectral intensity of $\gamma$-rays emitted from the jets, can be obtained 
from the following expression
\begin{equation}
I_{\gamma}(E_{\gamma})= \int_V Q_{\gamma} (E_{\gamma},z)d^3r = \pi tan \xi^2 \int_{z_0}^{z_{max}} Q_{\gamma} (E_{\gamma},z)z^2dz,
\end{equation}
where $F_{\gamma}$ is the spectrum of the produced $\gamma$-rays \cite{Kelner} with 
energy $x=E_{\gamma}/E_p$ for a primary proton energy $E_p$.

\section{Results and discussion}

Figure 1 shows the cooling rates of protons, pions and muons for the extragalactic 
binary system M33 X-7 (up) at the base of the jets. For comparison, the corresponding
results of the Galactic binary system SS 433 (bottom). The plots of the cooling rates 
of protons  show the synchrotron emission (solid lines), the adiabatic cooling (dotted lines), and the p-p collision (dashed lines). The plots for pions show respectively 
the synchrotron emission (solid lines), the pion-proton collision (dashed lines), 
the adiabatic cooling (dotted lines), and the decay rates of pions (dot-dashed lines).
Finally the plots of the cooling rates of muons show the synchrotron emission (solid
lines), the adiabatic cooling (dotted lines), and the decay rates of muons 
(dot-dashed lines).

\begin{figure}[ht] 
   \includegraphics[width=0.33\linewidth]{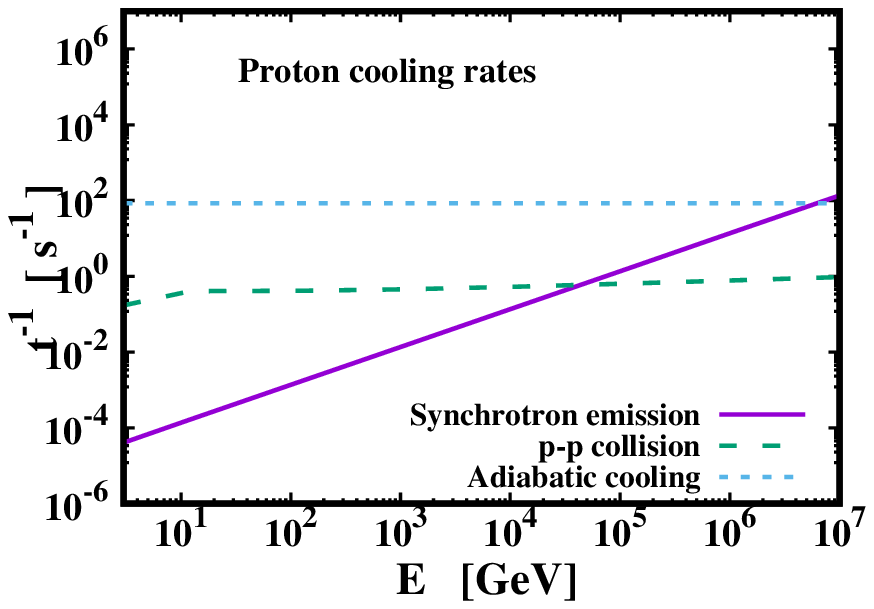}
   \hspace*{-0.5 cm} 
   \includegraphics[width=0.33\linewidth]{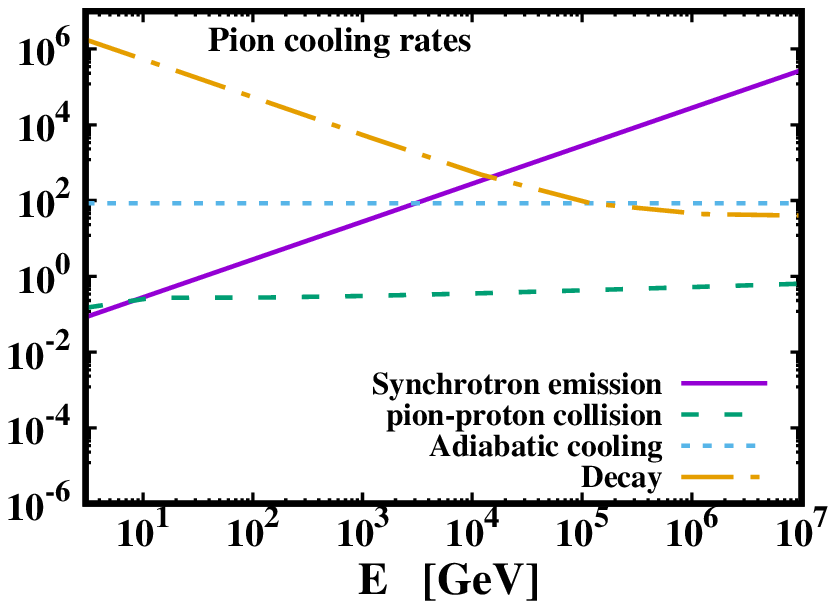}
   \hspace*{-0.5 cm} 
   \includegraphics[width=0.33\linewidth]{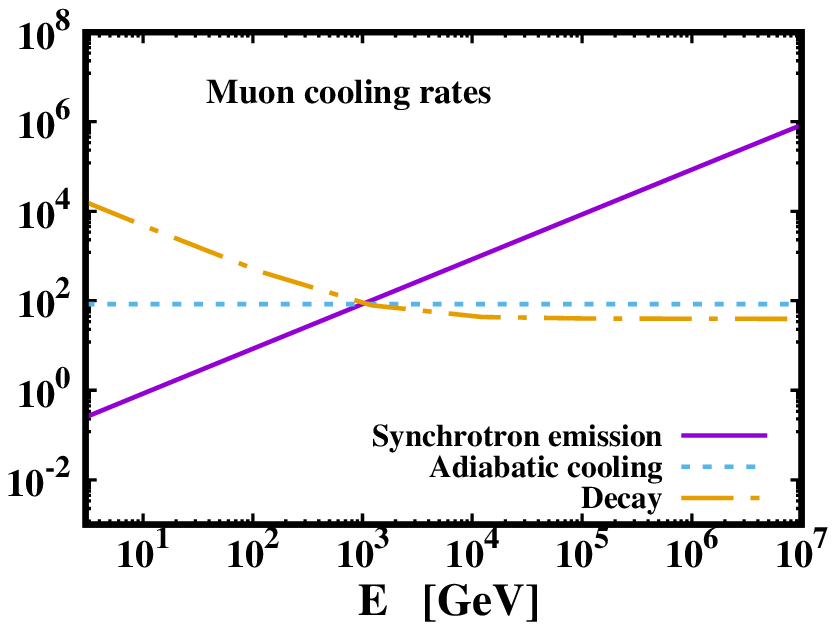}
   \\
   \includegraphics[width=0.33\linewidth]{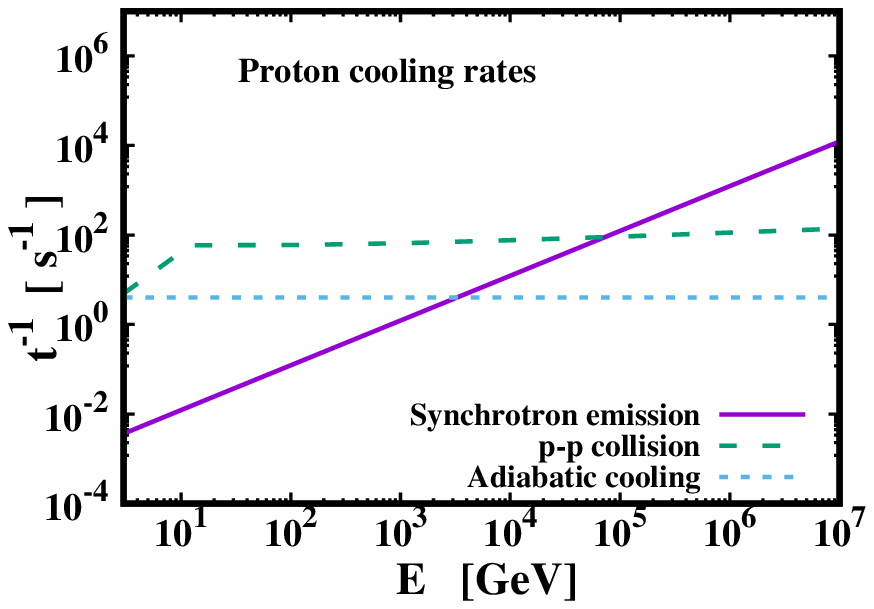}
   \hspace*{-0.5 cm}
   \includegraphics[width=0.33\linewidth]{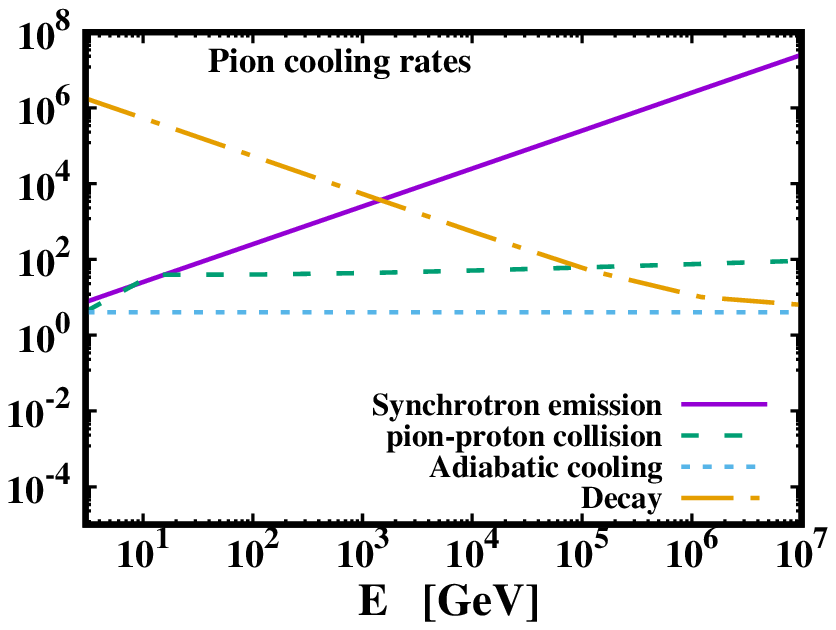}
   \hspace*{-0.5 cm}
   \includegraphics[width=0.33\linewidth]{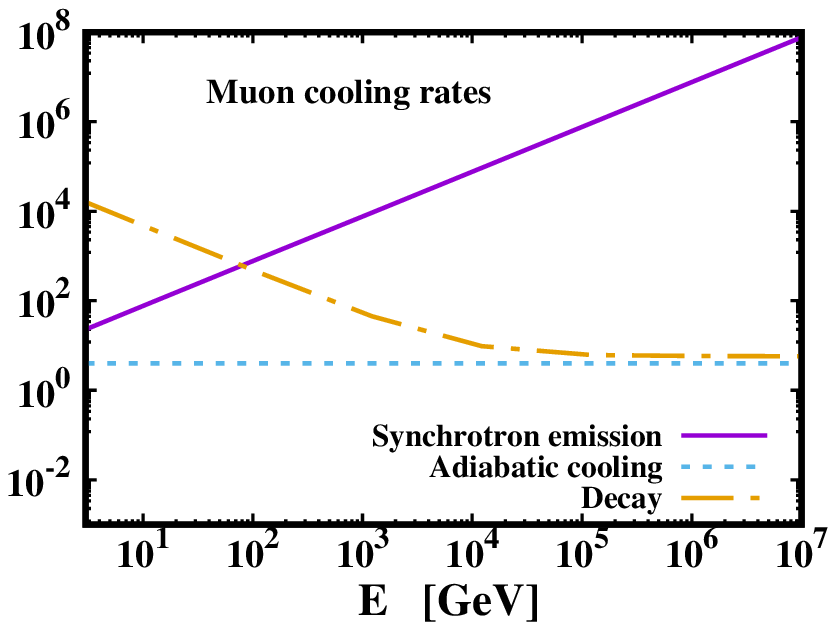}
   \vspace*{0.2 cm}
    \caption{Cooling rates for protons, pions and muons for M33 X-7 (top) and SS433 (bottom) at the base of the jets}
\end{figure}

From Fig. 1 it becomes obvious that , the particle synchrotron losses dominate the high energy region. 
In the case of protons, due to their mass, synchrotron losses are not dominant up to 
very high energies. For pions and muons, the decay losses dominate for lower energies, 
due to their decay rates. The main difference for these two MQs systems concerns the 
synchrotron cooling rates. For wide half-opening angles (M33 X-7, $\xi = 7^{\circ}$), 
the magnetic energy density is lower than that for narrow ones (SS433, $\xi=0.6^{\circ}$),
and, hence, the magnetic field is also lower. This leads to a lower synchrotron loss 
rate for the M33 X-7 system.

After obtaining the distributions for protons, pions and muons, the neutrino and 
$\gamma$-ray intensities may be calculated. Extensive results for $\gamma$-ray and 
neutrino production in various half-opening angles $\xi$ of the M33 X-7 will be
discussed elsewhere.

\section{Summary and Conclusions}

In this work we address radiative and neutrino emission from microquasars and X-ray 
binary stars (XRBs). These consist of a compact object (a stellar mass black hole or a 
neutron star) and a donor star. The strong gravitational field of the compact object 
is attracting the companion's star mass and an accretion disc in the equatorial 
region is formed which emits radiation and produces relativistic plasma jets 
all along the axis of rotation of the black hole. The neutrino and $\gamma-ray$ 
emissions originate from decay and scattering processes of the secondary particles 
produced by the $p-p$ scattering mechanism, i.e. the collision of relativistic 
protons of the jet with the thermal ones.

In this work we performed detailed calculations for the cooling rates of various
processes taking place in hadronic jets of the extragalactic system M33 X-7. 
These cooling rates enter the proton, pion and muon distributions through which 
one obtains neutrino and $\gamma-ray$ intensities. The results obtained depend 
crucially on the half-opening angle $\xi$ of the employed model. 

\section{Acknowledgments}
TSK acknowledges that this research is co-financed by Greece and the European Union
(European Social Fund-ESF) through the Operational Programme "Human Resources
Development, Education and Lifelong Learning 2014- 2020" in the context of the project
MIS-5047635.

\end{document}